\documentclass[11pt,twoside]{article}

\usepackage{asp2006}
\usepackage{epsf}
\usepackage{psfig}
\usepackage{lscape}

\newcommand{\U}[1]{\ensuremath{\mathrm{~#1}}}
\newcommand{\Myr}{\U{Myr}}

\newcommand{\kpc}{\U{kpc}}
\newcommand{\msun}{\U{M}_{\odot}}

\begin{document}
\title{Substructures formation induced by gravitational tides?}
\author{F.~Renaud\altaffilmark{1,2}, Ch.~Theis\altaffilmark{2}, T.~Naab\altaffilmark{3} and C.~M.~Boily\altaffilmark{1}}
\altaffiltext{1}{Observatoire astronomique and CNRS UMR 7550, Universit\'e de Strasbourg, 11 rue de l'Universit\'e, F-67000 Strasbourg, France}
\altaffiltext{2}{Institut f\"ur Astronomie der Univ. Wien, T\"urkenschanzstr. 17, A-1180 Vienna, Austria}
\altaffiltext{3}{University Observatory, Scheinerstr. 1, D-81679 Munich, Germany}

\begin{abstract}
Physics lectures always refer to the tides as a disruptive effect. However, tides can also be compressive. When the potential of two galaxies overlap, as happens during a merger, fully compressive tides can develop and have a strong impact on the dynamics of substructures such as star clusters or tidal dwarf galaxies. Using $N$-body simulations of a large set of mergers, we noticed the importance of these tidal modes at cluster scale. With a model of the Antennae galaxies, we conclude that the positions and timescales of these tidal modes match the actual distribution of young clusters. A detailed study of the statistics of the compressive tides shows a stunning correlation between this purely gravitational effect and the observed properties of the star clusters. In this contribution, we introduce the concept of compressive tide and show its relevance in the special case of the Antennae galaxies. We extend our conclusions to a broad range of parameters and discuss their implications on several critical points such as the infant mortality, multiple star formation epochs in clusters or the age distribution.
\end{abstract}

\section{Introduction}
The role of the tides in the evolution of galaxies has been known to be of prime importance since the earliest observations and simulations of such systems \citep[see e.g.][]{Sandage1961, Toomre1972, Toomre1977, Hibbard1996, Laine2003, Elmegreen2007}. With increasing resolution in both methods, detailed investigations on the structures formed in galactic pairs became possible. Among them, the study of star clusters and tidal dwarf galaxies (TDGs) gathered a lot of efforts with the closest major merger as a reference: the Antennae galaxies \citep[see e.g.][]{Barnes1988, Mirabel1992, Whitmore1995, Meurer1995, Hibbard2001, Fall2005, Mengel2005, Whitmore2007, Karl2008}. Recent HST observations of this pair revealed an amazing firework of star formation in the central regions. 

Though established, the link between the parsec-scaled structures like the young star clusters, and the phenomena covering over tens of kiloparsec, as tides do, still misses a quantitative description. We propose to partially bridge the gap between these two scales. \Citet{Renaud2008} noted that \emph{fully compressive} tides develop in interacting galaxies and have characteristic timescales of some $\sim 10 \Myr$, matching well those of the formation and early life of star clusters. They proposed that this additional gravitational effect would increase the binding energy of a young cluster and thus provide a favorable environment for the formation of stars.

Considering these tidal modes as the boundary conditions of a cluster, we propose that they help to form and to bind a cluster, and this way, prevent it from early dissolution \citep{Whitmore2007, deGrijs2007}. In this contribution, we extend the conclusions of \citet{Renaud2008} thanks to a parameter survey. After a brief description of the method used, we present the results of the Antennae galaxies and then, the influence of the pericenter distance between the progenitors on the statistics of compressive tidal mode. We close with a discussion on the role of the compressive tides on the evolution of star clusters.

\section{Numerical method}
The properties of the tides rely on the shape of the gravitational potential. To derive them, we use collisionless $N$-body simulations of typical galaxy mergers\footnote{The gas, which represents a small fraction of the mass, is here supposed to follow the stellar component and thus will not affect strongly the gravitational potential.}. The resolution used for this study counts 1,400,000 particles shared among the disk-bulge-halo components of the models. The smoothing length and the number of particles match the typical cluster scale ($40 \U{pc}$ and $10^5 \msun$).

The equations of motion are integrated by {\tt gyrfalcON} \citep{Dehnen2002}. The tidal tensor (i.e. the second spatial derivative of the potential, see \citealt{Renaud2008} for details) is computed at the position of a stellar particle and then diagonalized. The eigenvalues determine the strength of the tides along the associated eigenvectors. The nature of the tidal field only depends on the sign of the largest eigenvalue: when positive, the tides tend to stretch the matter along at least one direction; when negative, all the eigenvalues denote a compression and the tides are \emph{fully} compressive.

By applying this method via efficient algorithms to all the stellar particles, we derived statistics of the compressive tidal mode, and its evolution in time. Details of these methods will be presented in a forthcoming paper \citep{Renaud2009}.

\section{The Antennae}
Our starting point is the model of the Antennae galaxies presented in \citet{Renaud2008}. The two self-gravitating equal-mass progenitors have been set up with {\tt magalie} \citep{Boily2001} which gathers an exponential disk, a \citet{Hernquist1990} bulge and an isothermal dark matter halo, all scaled to get two S0 galaxies. The progenitor have been set on a bound orbit with an eccentricity $e \approx 0.96$ and an initial separation of $75 \kpc$.

\begin{figure}
\plotone{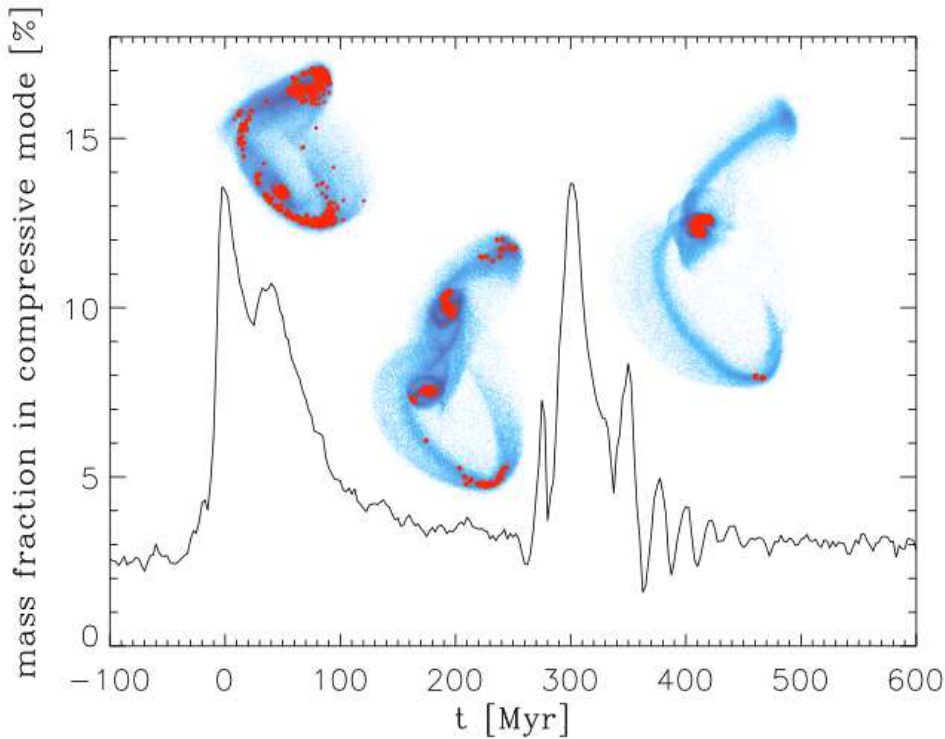}
\caption{Evolution of the mass fraction in compressive mode for our Antennae model. The equivalent morphology is shown for three snapshots ($50 \Myr$, $150 \Myr$ and $300 \Myr$) with column density (in blue) and the compressive particles (in red). The two pericenter passages ($t = 0$ and $t = 300 \Myr$) are associated with the two main peaks, rising the mass fraction in compressive mode to $\approx 14\%$.}
\label{fig:ant}
\end{figure}

In Fig.~\ref{fig:ant}, we show the evolution of the stellar mass fraction in compressive mode. Before the first passage ($t < 0$), the galaxies are isolated and have an intrinsic compressive fraction of $\sim 3 \%$. The main tidal events like the creation of bridges and tails ($0 < t < 100 \Myr$) and the merger phase ($250 \Myr < t < 450 \Myr$), are clearly visible on this plot, as they are linked to a raise of the mass fraction by a factor $\sim 5$. Note that the compressive zones spread in the nuclei area but also in the tails, and in particular near the TDGs region \citep{Schweizer1978, Mirabel1992, Hibbard2005}.

Typically, one of our stellar particle enters and leaves compressive regions and thus undergoes a series of on/off episodes along its orbit. The age distribution of these events follows an exponential law with a timescale of $\sim 10 \Myr$. By considering a mean distribution between continuously compressive events and on/off episodes, it is possible to retrieve a power-law $dN/d\tau \propto \tau^{-1.1}$ age distribution of the compressive modes. With the rough assumption that star clusters would dissolve when the tidal field switches from compressive to extensive, one can directly compare this distribution with the observations of \citet[$dN/d\tau \propto \tau^{-1}$]{Fall2005} for the age of the Antennae clusters.

Because of a growing number of clues suggesting a link between compressive tides and young clusters, we decided to extend our conclusions to a larger set of parameters. Among them, we focus here on the influence of the pericenter distance.

\section{Influence of the pericenter distance}
\begin{figure}
\plotone{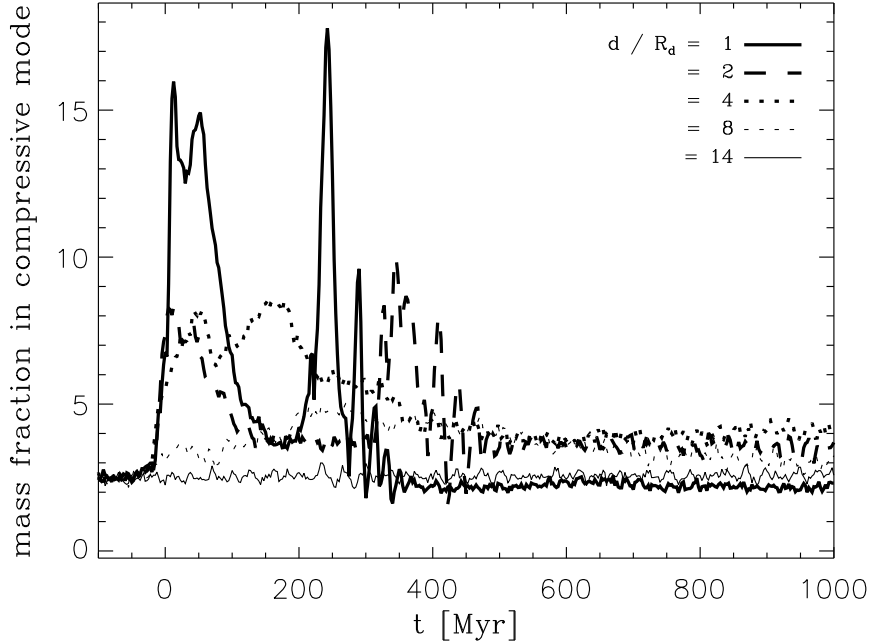}
\caption{Evolution of the mass fraction in compressive mode for different pericenter distances $d$, normalized to the exponential disk scalelength $R_d$ of the progenitors. The fraction increases at the pericenter passage(s) when the interaction is penetrating. In the other cases, it remains almost flat, at the level of the isolated progenitors ($\sim 3\%$).}
\label{fig:dist}
\end{figure}

In this section, we investigate the role of the distance between the two progenitors at the first passage. For simplicity, all the other parameters (scalelengths, inclination, spin, eccentricity ...) remain as in the Antennae model. We set the orbits of the galaxies to cover the full range of distances, from mergers (short distance at first passage) to fly-bys. The Fig.~\ref{fig:dist} plots the evolution of the mass fraction in compressive mode (same as Fig.~\ref{fig:ant}) for different pericenter distances $d$, normalized to the scalelength of the exponential disk $R_d$ of the progenitors.

Again, the main steps of the history of the interaction are clearly visible. However, we note that the mass fraction in compressive mode decreases as the pericenter distance grows. As for the Antennae galaxies ($d / R_d = 1.3$, not shown here), when one progenitor goes deeply into the potential well of the other ($d / R_d = 2, 4$), the fraction in compressive mode increases by a factor four to five. On the contrary, wider interactions show a milder increase ($d / R_d = 8$) or remain at the level of isolation ($d / R_d = 14$). In this fashion, if the compressive tidal modes are to be linked to an enhanced star formation rate (SFR), we would expect a higher rate for close passages, thus mergers.

Other studies showed an enhanced SFR, even for fly-bys \citep[see e.g.][]{diMatteo2007}. This difference may be interpreted as gas stripped off one galaxy to the other, a phenomenon that occurs for all types of interactions. Yet, it is difficult to predict an accurate fate of the giant molecular clouds or the young star clusters, as their response to a compressive non-isotropic non-constant tidal field has not been investigated so far.

\section{Compressive tides and multiple populations}
The statistical study of the tidal field revealed global properties that have been extended to the general case of galaxy mergers \citep{Renaud2008, Renaud2009}. However, the actual impact of the tides on a single cluster-sized element may vary from the statistics and is generally much more complex. As an example, let's consider one of our stellar particle, orbiting in the merger. It will likely experience a series of on-off compressive episodes, with different periods. On a speculative note, it is possible to imagine that external gas and dust would be driven by the compressive modes into the cluster core and then, condensed to form new stars. If the compressive events are well separated in time (and thus, in space along the orbit of the cluster), the chemical properties of this additional gas may differ from those of the original interstellar medium that formed the first generation of stars. In this fashion, it is possible to explain multiple populations, with different helium abundances, as observed in some resolved massive clusters \citep[see e.g.][]{Piotto2007, Milone2009}. Furthermore, it is possible that slow stellar ejecta would be kept bound by the additional gravitational effect of the compressive tides, and then be processed into one or more secondary generations of stars. This would favor a self-enrichment of the cluster with respect to a tidally unperturbed system. Yet, a quantitative study of the impact of the compressive modes on the hydrodynamics of a star cluster is at a too early stage and this point remains speculative.

\section{Conclusions}
In this contribution, we showed that fully compressive tidal modes develop in the course of a prototypical merger like the Antennae galaxies. Accounting for $\sim 15\%$ of the stellar mass, they last long enough ($\sim 10 \Myr$) to prevent the newly formed cluster from early dissolution. A good correlation has been found between the position of the compressive regions and the observed substructures (star clusters and TDGs). When trying to extend these conclusions to a larger set of parameters, we showed that the fly-bys would not be strongly affected by the compressive modes. Indeed only mergers present a significant increase in the mass fraction entering such modes, matching the epochs of the orbital passages. Furthermore, the amplitude of these increases is anti-correlated with the pericenter distance: short distances induced a large rise in the mass fraction in compressive mode, while wide interactions are closer to the quiescent stage. Typical proto-clusters also show an on-off behavior of the compressive tides. This might give rise to multiple star formation epochs, with or without self-enrichment of the cluster. However, the exact response of parsec-scale structures embedded in a tidal field varying in time and direction has still to be addressed and will be presented in a forthcoming paper.

\acknowledgements
FR is a member of the IK~I033-N \emph{Cosmic Matter Circuit} at the University of Vienna. This project has been supported by the DFG Priority Program 1177 `Galaxy Evolution'.

\end{document}